\documentclass[12pt]{article}
\usepackage{a4wide,epsfig,psfig,amsfonts,latexsym}

\newcommand{\abs}[1]{\left| \, #1 \, \right| }
\newcommand{\skl}[1]{\left\langle \, #1 \, \right\rangle }

\begin{document}

\title{\bf A simple model of epitaxial growth} 

\author {  {\sc M.Biehl, W. Kinzel,  and S. Schinzer}\\
            Institut f\"ur Theoretische Physik, \\
            Julius--Maximilians--Universit\"at W\"urzburg\\
            Am Hubland, D--97074 W\"urzburg, Germany\\ }

\date{}

\maketitle
 \mbox{~~~~~~}PACS.  81.10.Aj, 81.15.Hi, 05.70.Ln, 68.55.-a \\

\begin{abstract}
 A discrete solid--on--solid model of epitaxial growth
is introduced which, in a simple manner, takes into account the
effect of an Ehrlich--Schwoebel barrier at step edges as well as the
local relaxation of incoming particles. Furthermore a fast step edge 
diffusion is included  in $2+1$ dimensions.  The model exhibits the
formation of pyramid--like structures with a well--defined constant
inclination angle. Two regimes can be distinguished clearly:  in an initial phase (I)
a definite slope is selected while the number of pyramids remains unchanged.
Then a coarsening process (II) is observed 
 which decreases the number of islands according to a power law in time.
Simulations support self--affine scaling of the growing surface in both regimes.
The roughness exponent is $\alpha =1$ in all cases. For growth in
 $1+1$ dimensions we obtain dynamic exponents $ z = 2 $ (I) and $ z = 3 $ (II).
Simulations for $d=2$ seem to be consistent with  $z= 2$ (I) and
 $z= 2.3$ (II) respectively.  
\end{abstract}

\ \\

Among the different morphologies of growing surfaces in
molecular beam epitaxy, the frequently observed formation and coarsening 
of pyramid--like structures with a well--defined slope is of particular
interest \cite{sp94,bar95,sp96,lk97,kru97b}. This kind of behavior has
been seen in experiments of molecular beam epitaxy in such
diverse systems as  Fe(001) \cite{tkw95,sps95}, Cu(001) \cite{zw97},
GaAs(001) \cite{joh94, ojl95}, and HgTe(001) \cite{oes96}. 
The selection of a constant inclination angle can be associated with the
compensation of slope--dependent uphill and downhill currents of
deposited particles. This has been studied by numerical integration of
continuum equations \cite{sp94,bar95,lk97}.

In the framework of our discrete model, particles are deposited on a
$d$--dimensional simple cubic lattice with $L^d$ sites.  The
prescription obeys a solid--on--solid restriction, i.e.\ neither
overhangs nor holes will occur and the growing structure is completely
characterized by integer height variables assigned to each of the
sites.  The time $t$ associated with the growth process is defined as
$ t = N / L^d $ where $N$ is the total number of deposited particles.
We will consider a relaxation mechanism for the incoming particles as well as 
an additional diffusion process with high energy barriers at step edges.  
Only after the particle has reached its final position, a new one is deposited
on the surface.

To begin with, the model is explained in terms of a one--dimensional
substrate ($d=1$) which is initially flat ($h_j (t=0) = 0$ for $ j =
1,2,\ldots, L $) and satisfies periodic boundary conditions.
A new particle is placed with equal probability anywhere on the
surface, say at site $i$.  If any of the neighbor sites $j$~ (with $
\left|j-i\right| \leq R_{inc}$) offer a lower height $ h_j < h_i$, the
freshly deposited particle is moved to the lowest of these sites (a random
one in case of a tie). 
This  models the effect of particles reaching the
substrate with a residual momentum and kinetic energy \cite{est90}.
The parameter $R_{inc}$ is called the {\sl incorporation radius\/} of the 
model.

Without additional diffusion this prescription would coincide with the
well--known {\sl Random Deposition with Surface Relaxation\/} (RDSR)
\cite{bar95}. 
  RDSR belongs to the
Edwards--Wilkinson universality class the scaling behavior of which is known
analytically \cite{bar95,ew82,nt92}. In particular,  the surface becomes flat 
in $2+1$ dimensions, i.e. its width
increases only logarithmically with time and system size. 

However, if the incoming particle is not yet bound to in--plane neighbors
after relaxation, it is considered to  diffuse along the
surface.  Now, an energy barrier at step edges can effectively
surpress downward moves since the above mentioned residual momentum has been 
lost.  In the following, this Ehrlich--Schwoebel barrier
is taken to be infinite.
Consequently, the
diffusion is restricted to connected sites of equal height $h_i$.
The main  result of this Letter is that the interplay of
this diffusion and the incorporation mechanism yields  a stable
slope which is controlled by  $R_{inc}$ only.

The second  parameter of the model is the diffusion length
$l_D$  which marks the maximum distance that the particle can explore
before it comes to rest.  A simplistic prescription is used to replace
the actual diffusion in $1+1$ dimensions: if a potential in--plane
neighbor    
is available within a distance $l_D$ the particle is placed there, otherwise
it sticks to the deposition site.  This mimics the nucleation of
diffusing particles with a mean free path of order $l_D$ \cite{pvw92}, a quantity
which depends on the ratio of the diffusion constant and
the incoming flux (see e.g.\ \cite{vpt92}).
 
 The diffusion process alone (without relaxation) does not alter the
 frequencies at which values $h_j$ occur in comparison with simple
 {\sl Random Deposition\/} (RD). Thus, formally, $ \skl{h} =t$ and
 $w(t) \,= \, (\skl{h^2} -
 \skl{h}^2)^{1/2} \, \propto t^{\, \beta} \mbox{~~with~~} \beta =1/2$
 where $\skl{\ldots}$ denotes the average
 over all lattice sites.  The width $w$ of the surface grows
 indefinitely and no saturation due to lateral correlations is
 observed.  However, the spatial distribution of $h$--values is
 changed drastically. The infinite Ehrlich--Schwoebel barrier results in the
 formation of separate {\sl islands \/} of linear size $\propto l_D$
 on the flat substrate. Without local relaxation no coarsening is
 possible, the number of islands and their sizes remain constant
 after a short transient. However, their height and thus the slopes
 grow indefinitely resulting in the RD power law. The profile
 of a single island is given by the inverse of a cumulative binomial
 distribution (i.e.\ an inverse error function in the continuum limit)
 as was observed independently by Krug \cite{kru97a}.

\begin{figure}[t]
\begin{center}
\setlength{\unitlength}{0.8pt}
\begin{picture}(325,150)(0,0)
 \put(0,0){\makebox(325,150)
          {\includegraphics{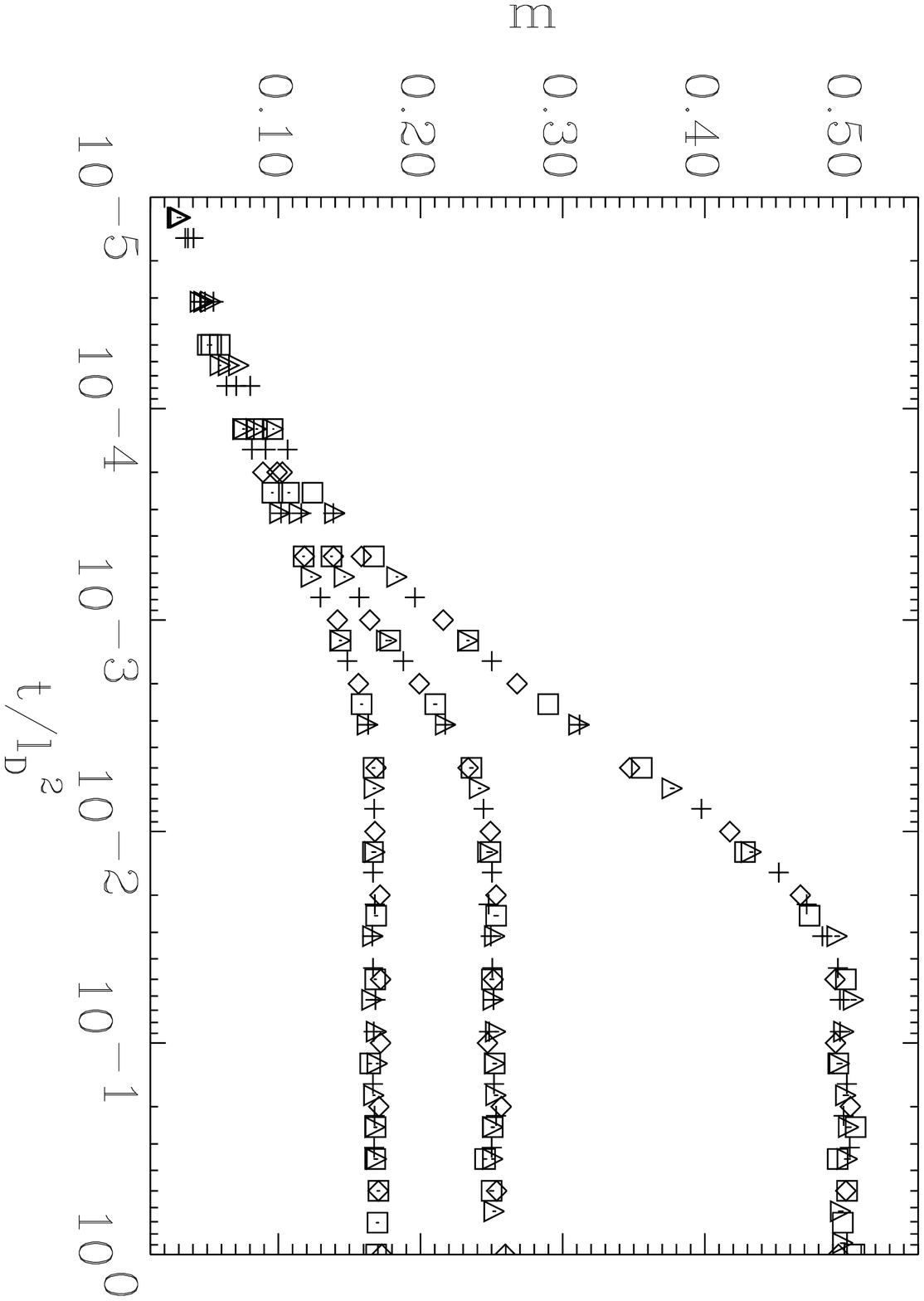}}}
 \put(0,0){\makebox(325,150)
          {\includegraphics{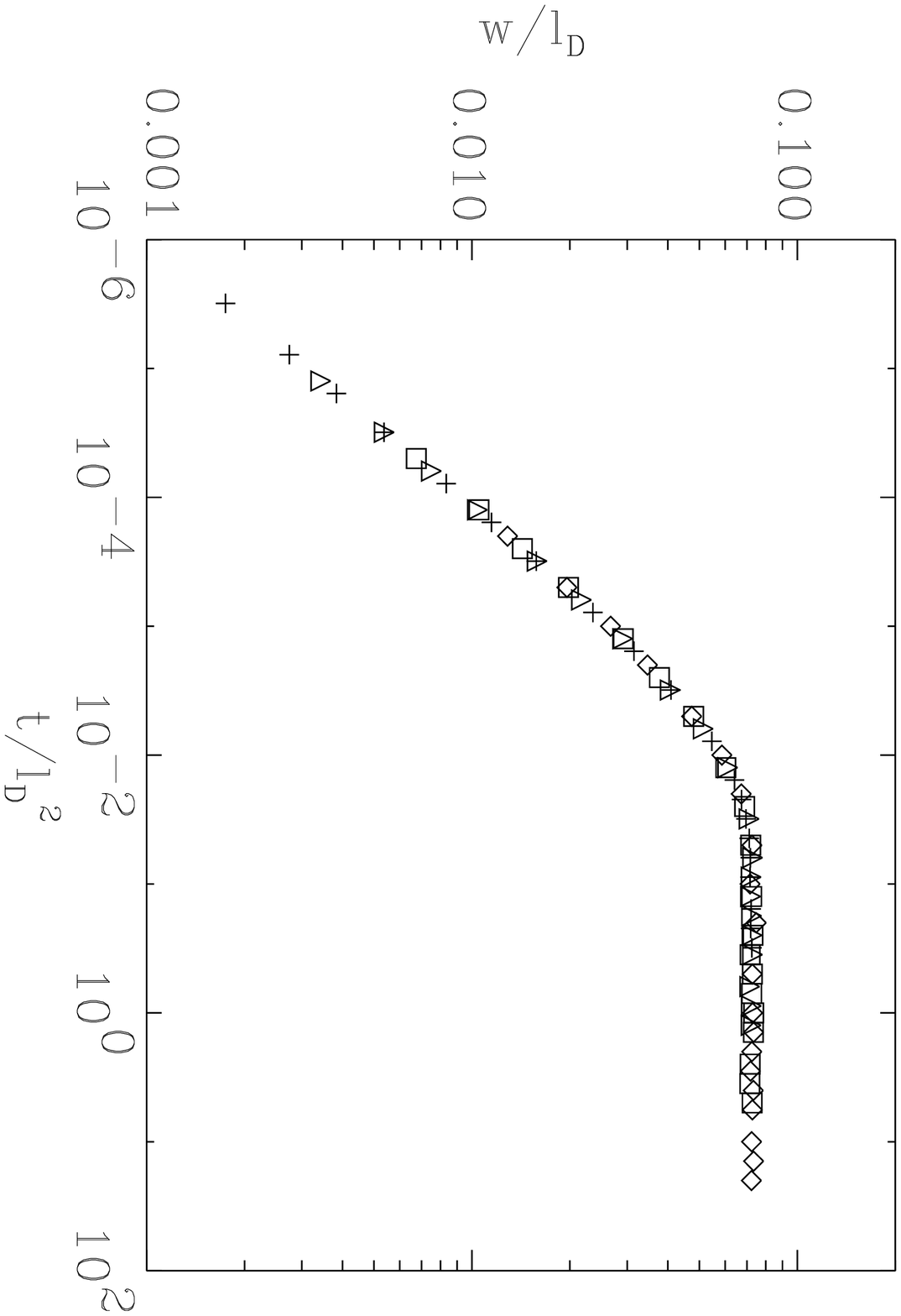}}}
\put(0,0){\makebox(-60,260){\bf a)}}
\put(0,0){\makebox(470,260){\bf b)}}
\end{picture}                    
\end{center}
\caption{Growth in $1+1$ dimensions: evolution of a single island for various system sizes 
$ L= l_D = \, 100 (\diamond)$, $ 200 (\Box)$, $400 (\triangle)$, and $800 (+)$.
Results were averaged over $50$ independent runs, error bars would be smaller than
the symbols.  
a) shows the average local slope $m$  vs. scaled time for three different values of the incorporation radius
(top to bottom: $ R_{inc}=1,2,3$). 
In b) the evolution of the surface width $w$ is displayed for the same system sizes
and $R_{inc} = 1$.  }
\label{figure1}
\end{figure}

 The net effect of the infinite Ehrlich--Schwoebel barrier is an
uphill current of deposited particles which can be compensated for by
the relaxation process described above.  The resulting slope selection
is most clearly studied for  $l_D = L$ when only one island
forms.
Figure 1(a) shows for this case  the evolution of the average local height 
difference $m = \skl{\abs{h_{j+1}-h_j}} $
for three different values of $R_{inc}$. (The profiles
produced by our model are strictly monotonic along the slopes of the pyramids, otherwise $m$ would 
systematically over--estimate their inclination.) 
The curves for different system sizes $L=l_D$ collapse when time is scaled 
with $l_D^{\, 2} $, indicating that after a characteristic
time $t_m \, \propto l_D^{\, 2} $ a constant slope is reached.

Note that if a particle is deposited on a terrace of length $n$, it jumps
downward with probablity $R_{inc} / n$   whereas it is added to the upper 
terrace with probability $1- R_{inc} /n$. Therefore,  the selected slope 
for which uphill and downhill currents cancel is $ m_{sat} = (2\, R_{inc})^{-1} $ 
in excellent agreement with the simulations.
A more detailed presentation of this argument, also in terms of a continuous
description of the step dynamics, will be given in a forthcoming publication
\cite{sk97}.

While the single island  forms, the width $w$ of the corresponding 
surface  increases until it reaches its saturation value  $ w_{sat} \, =
 m_{sat} \, L \, / \sqrt{48} $. Thus, the roughness exponent is $\alpha = 1$ 
as expected for  a constant slope. 
Simulations suggest a growth exponent
 $\beta =1/2 $  which corresponds to a dynamic exponent $z=2$ in agreement with the
time scale found for the slope selection process.  Apparently, 
the relaxation process is basically irrelevant for the small time behavior
and $w$ increases as in Random Deposition.
Figure 1(b) displays the evolution of $w$ for various system sizes, rescaled
according to the assumed behavior.  

Next we investigate the dynamics of many islands. For $ 1 <\!\!< \, l_D \, <\!\!< L$
the number of initially formed pyramids is of the order of $L/l_D$. The small time behavior is
dominated by the slope selection process which is completed after the characteristic
time  $t_m \propto  l_D^{\, 2}$.
In the second phase fluctuations in the deposition of particles induce a 
coarsening process in which smaller islands are merged into larger ones \cite{tsv97}.  The
number of islands  $n_{isl}$ 
decreases according to  a power law:  ~$ n_{isl} \propto t^{- 1/3 } $.
This  exponent coincides with theoretical predictions for noise driven 
coarsening  in one--dimensional systems \cite{sp96,tsv97} and can be 
related to domain growth in magnetic systems \cite{ls61,wag61,km85}.
Accordingly, the surface width increases like $ w \propto t^{1/3} $.  
After a time of order $L^3$  the system reaches its saturation state: 
a single remaining pyramid with the above given constant inclination. 

\begin{figure}[t]
\begin{center}
\setlength{\unitlength}{0.8pt}
\begin{picture}(325,150)(0,0)
 \put(0,0){\makebox(325,150)
          {\includegraphics{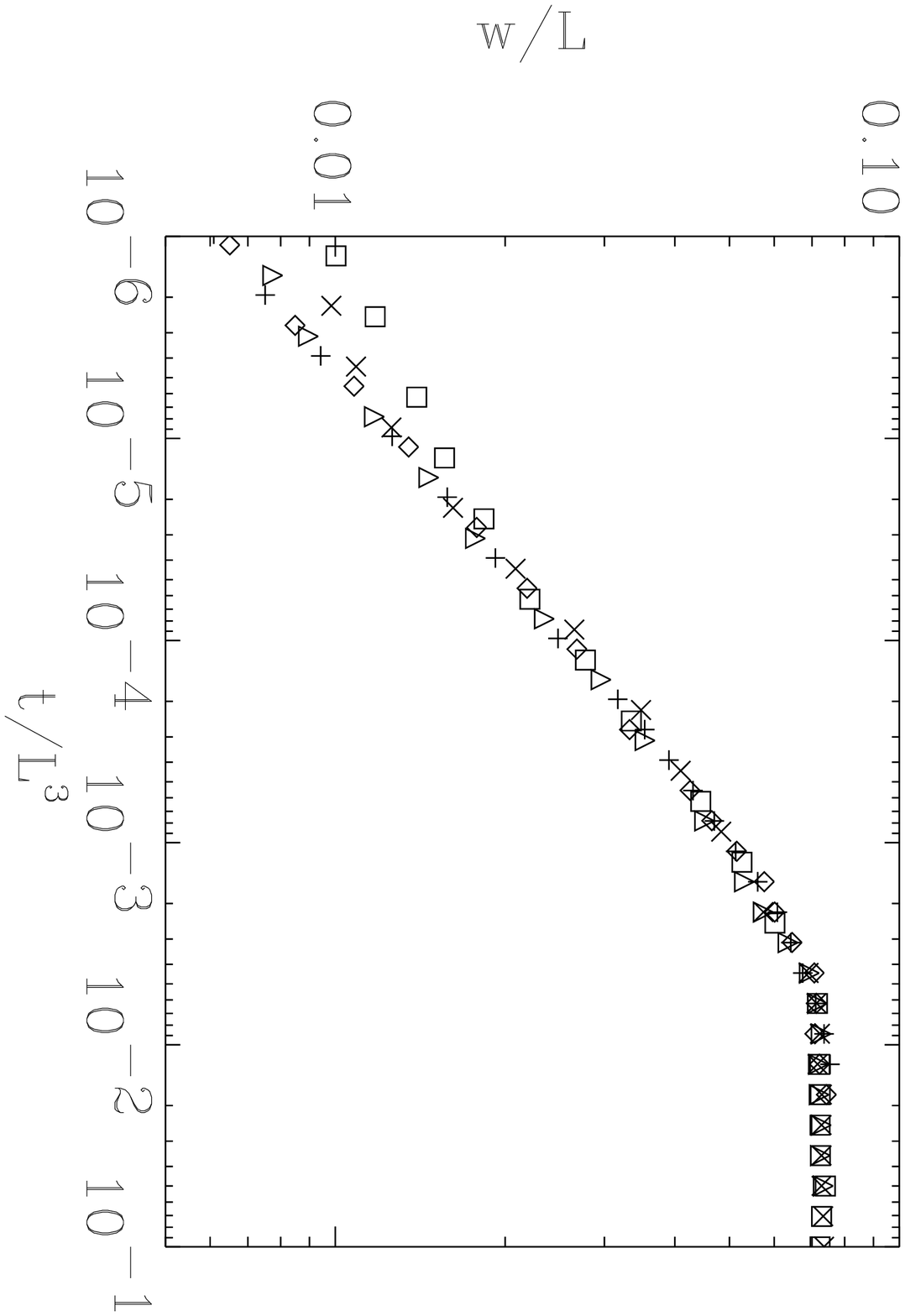}}}
 \put(0,0){\makebox(325,150)
          {\includegraphics{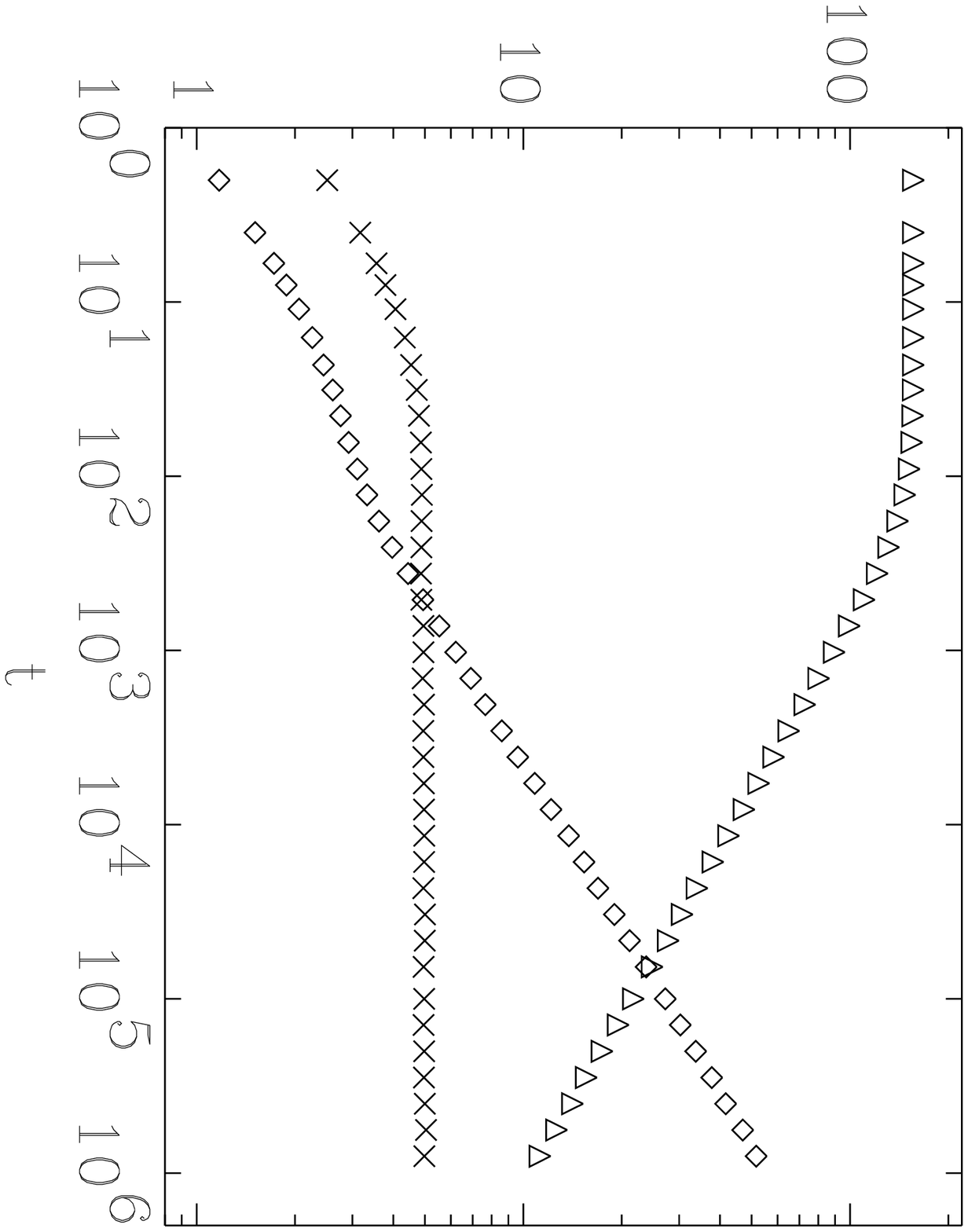}}}
\put(0,0){\makebox(-60,260){\bf a)}}
\put(0,0){\makebox(470,260){\bf b)}}
\end{picture}                    
\end{center}
\caption{The dynamics of the model for $ l_D =20 \ll L$ in $1+1$ dimensions: 
  a) surface width vs.\ time, rescaled according to the assumption
  $ z=3 $ and $\alpha =1$ for system sizes $ L = 200 (\Box)$, $ 283 (\times)$,
  $ 400  (\triangle )$, $566 (\Diamond)$,  and $ 800 (+)$.
   Averages were performed over $50$ runs with
  standard errors smaller than the symbol size.     
 In b) the evolution of  $w (\Diamond)$,  $10\cdot m (\times)$,
   and $n_{isl}$ $(\triangle)$ is shown  for
   a system with $L=5000$ and $l_D =20$ on average over $40$ simulations.
   After the slope is selected, surface width and number of islands evolve 
   according to $w \propto t^{1/3} $  and  $n_{isl} \propto t^{-1/3} $. }
\label{figure2}
\end{figure}

Fig 2(a) shows results of simulations with $l_D = 20$.  Under a
rescaling according to $z=3$ and $\alpha =1$ the curves for different
system sizes collapse for times greater than the $L$--independent
value $t_m$.  The above scaling law is very well confirmed by
Monte Carlo results provided the number of islands is still fairly
large. Fig 2(b) shows the results for $L=5000$ and $l_D =20$ with
$n_{isl}$ decreasing from about $150$ to $10$.  A more detailed
analysis indicates that, close to saturation, the pairwise merging of
the last few islands occurs at characteristic times and yields a
cascade--like increase of $w$.

It is important to note that the condition  $l_D \gg 1$ has to be satisfied in order
to produce the observed behavior. Otherwise, nucleation is likely to occur even  on small
terraces, the above mentioned monotonicity of the profile is lost, and  
$n_{isl}$  can fluctuate strongly in time. The detailed behavior of systems 
with a small diffusion length $ l_D = {\cal O} (1)$ will be the objective of 
forthcoming investigations. 

In the following we discuss a generalization of the model to 2+1 dimensions.
We assume  $R_{inc}=1$ and the diffusion process is simulated explicitely as a 
random walk under the constraint that the particle remains at the same height. 
The walk stops when the particle reaches an in--plane neighbor or 
has accomplished $n_D$  jumps.
This corresponds to a typical diffusion length $l_D \approx \sqrt{n_D}$
provided the set of accessible sites has a compact shape (unlike, for example, a
fractal cluster).

Hereupon the particle performs an additional diffusion (random walk at
constant height) along the step edge until it is bound 
to (at least) two in--plane neighbors or until $n_{D2}$ moves have been
accomplished. 
We would like to emphasize that without step edge diffusion the properties
of the model are changed drastically.  
For instance, we find a roughness exponent  $\alpha \approx 0.8$ for $n_{D2}=0$
whereas the expected slope selection with  $\alpha = 1$ is confirmed for $n_{D2} \gg 1$. 
The crossover is in itself interesting but here we concentrate on the model
with $l_{D2} \propto \sqrt{n_{D2}} $ large enough (of order $L$)
to ensure that the particle reaches the nearest accessible kink site.
In this regime the     
bases of the growing pyramids are virtually of square shape, cf.\ Fig. 3,
whereas smaller values of $l_{D2}$ yield rounded cross--sections.

The overall behavior is qualitatively the same as in 1+1
dimensions. First, a constant slope builds up which is observed
most clearly in the formation of a single pyramid ($\sqrt{n_{D}} \approx L$).
Figure 4(a) shows that here the surface width obeys a scaling relation
$ w = L^\alpha \cdot g(t/L^z)$ where  $ g(x\gg 1)=const. $
Again, slope selection occurs ($\alpha=1$). The systematic deviation  
from the prediction $m_{sat} = 1/2 $  is of order $1/L$ in finite systems, cf. inset in
 Fig.\ 4 (b).
This correction is taken into account in Fig.\ 4  by scaling $w$ with 
 $(m_{sat} L)$ rather than with $L$.

The observed dynamic exponent is in good agreement with  $z=2$. 
If $g(x)  \propto t^{\, \beta}$ for $x\ll 1$, this would imply
$\alpha / z  = \beta = 1/2$ as in $1+1$ dimensions. Yet, the simple power law seems 
to hold only for very short times.
This can be seen clearly in the inset of Fig. 4(a) where the
effective exponent obtained from two consecutive data points is plotted.
As already observed in 1+1 dimensions, the surface width increases  initially like
in RD with $\beta = 1/2$. 
The relevance of the incorporation mechanism is determined by the number of step edges  which 
is independent of the pyramid size in 1+1 dimensions.  However, for $d=2$  it increases 
with the island radius already in the submonolayer regime. Thus, the deviation from RD--behavior
becomes noticeable much earlier. 

\begin{figure}[t]
\begin{minipage}[t]{4.4cm}\makebox[0cm]{}\\
\psfig{file=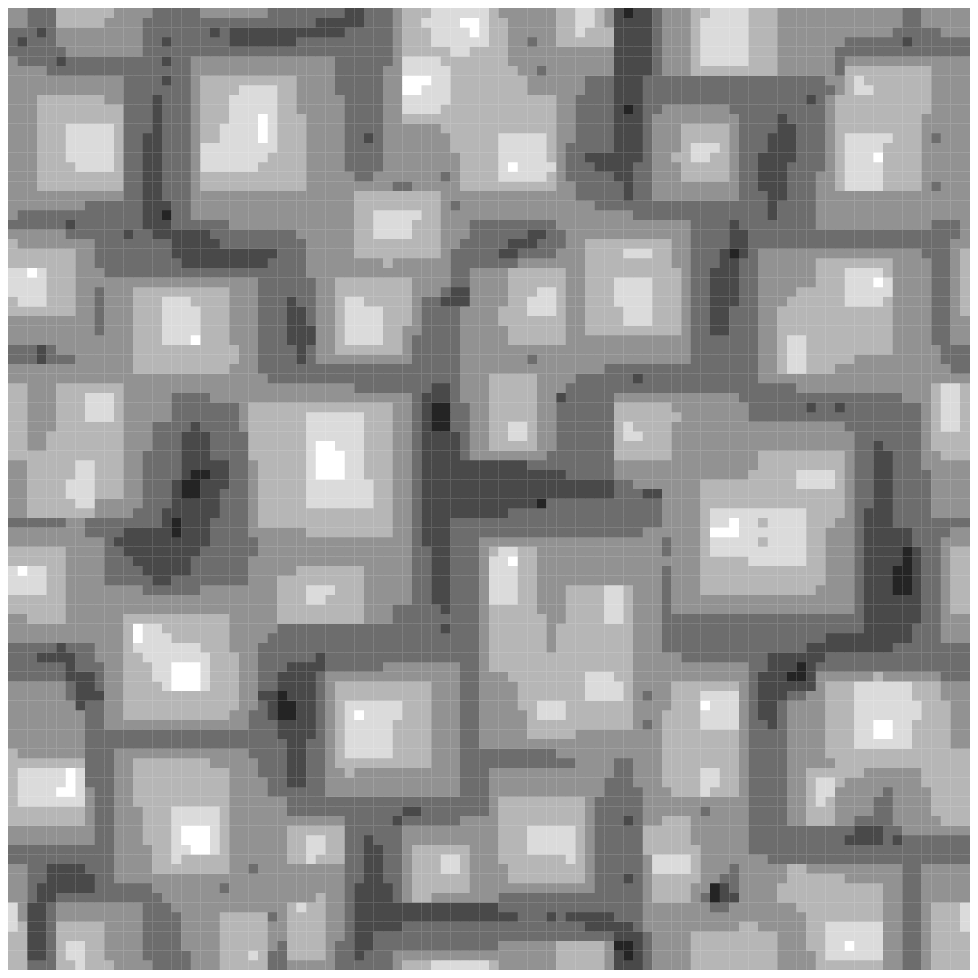,width=4.4cm}
\end{minipage} \hfill
\begin{minipage}[t]{4.5cm}\makebox[0cm]{}\\
\psfig{file=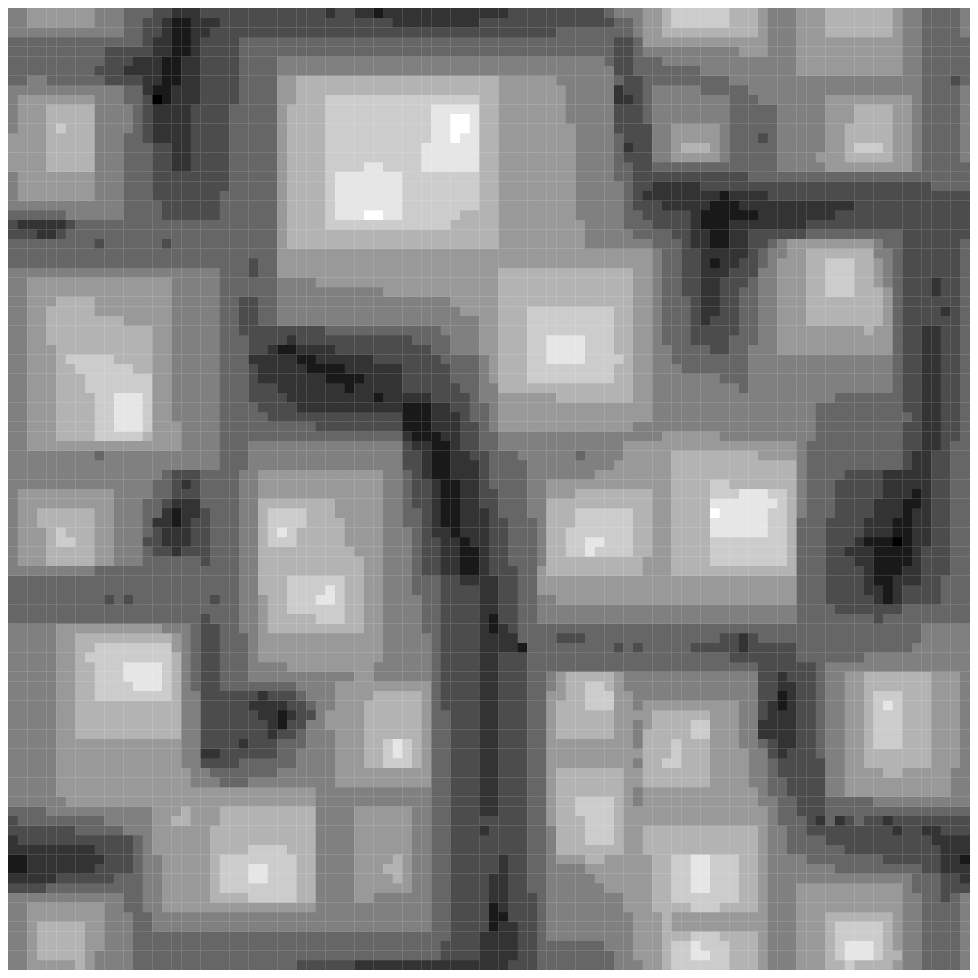,width=4.4cm}
\end{minipage} \hfill
\begin{minipage}[t]{4.4cm}\makebox[0cm]{}\\
\psfig{file=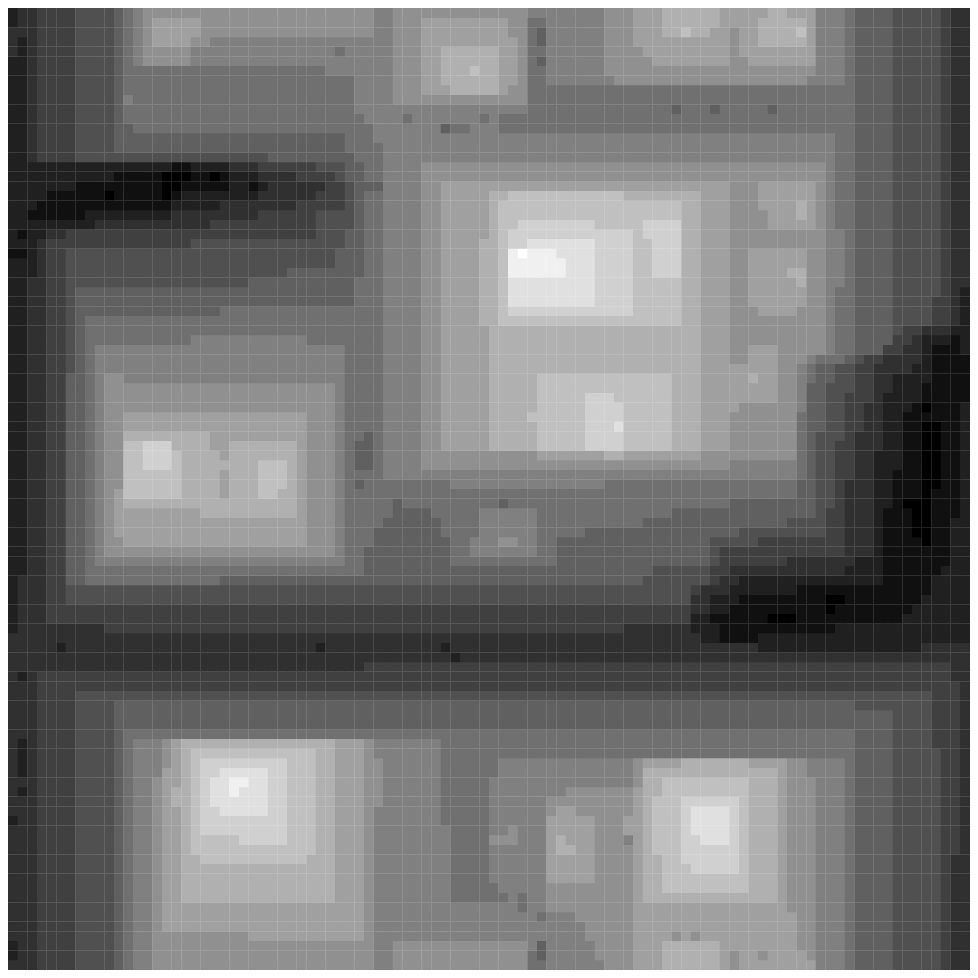,width=4.4cm}\\
\end{minipage} \hfill
\caption{Snapshots (density plots) of the evolution of the surface
morpholgy in $d=2$ at times $t=4,16,64$. The system size is $L=100$, the model
parameters are $R_{inc}=1$, $l_D=15$, and $l_{D2}=200$.}
\end{figure}

Preliminary results concerning the coarsening of many islands  ($l_{D} \ll L$)
are shown in Fig.\ 4(b). 
The data seem to be consistent with a  dynamic exponent
$z \approx 2.3  $  ($\beta \approx 0.43$) in this regime. 
Simulations with small and intermediate values of $n_{D2}$ indicate that $ \beta $
is  significantly smaller when the step edge diffusion is  slower.
However, a more precise determination
of this value is necessary and should be based on the simulation of larger systems. 
At the current stage a definite conclusion concerning the scaling behavior  cannot be drawn.
The above mentioned analogy to the dynamics of domains in magnetic systems would imply $1/\beta=z=3$
but is unlikely to carry over to epitaxial growth in 2+1  dimensions \cite{sp96}.
Several authors obtain  values $ \beta \leq 0.25 $  from the numerical studies of continuum equations 
 (see \cite{sp96} and refs.\  therein). A recent argument by Tang {\sl et al.\/} \cite{tsv97} 
suggests that $z = 2+d$ for noise--driven coarsening.

The relevance of our simple model for molecular beam epitaxy is
readily recognized if one identifies the parameter $l_D$ with the typical
island distance in the submonolayer regime. 
Several authors have calculated this length for the simple case
in  which two adatoms already form a stable nucleus (\cite{vpt92} and refs. therein).
For $d=2$ one obtains $l_D \propto (D/F)^6$ ($(D/F)^4$ in $d=1$ \cite{pvw92}) where $D$ is the
diffusion constant and $F$ the flux of arriving particles. 
A helpful simplification of our model is due to the infinite
Ehrlich--Schwoebel barrier. However, even with a finite (but non--vanishing) 
barrier  the qualitative scenario will be the same. 
Extensions of the BCF theory \cite{bcf51,ev94} which include an incorporation mechanism
also yield slope selection \cite{sts96}. There, however, the tops of pyramids are
rounded. Furthermore the terrace width depends on the actual energy barriers and the 
temperature of the system \cite{sk97}.

In summary, we have presented a simple model which reproduces essential features of
epitaxial growth. 
Local relaxation leads together with surface diffusion in the presence of a high
Ehrlich--Schwoebel barrier to the selection of a stable slope of growing islands. 
We identify two different scaling regimes (both with roughness exponent $\alpha =1$): 
the initial slope selection (I) and the coarsening of pyramids (II). 
For growth on a one--dimensional substrate we  find
 $z =2 $ in regime (I) and $z =3$ in (II).  Simulations in $d=2$ suggest the same dynamic
exponent for (I)  and  a value of  $z \approx 2.3$ in regime (II).
Apart from a more  accurate  measurement of these exponents in larger systems,
forthcoming investigations  will address the crossover behavior for  $l_D = {\cal O} (1)$, 
the precise influence of the step edge diffusion length $l_{D2}$, 
the effect of a finite Schwoebel barrier, and  a continuous description 
of the step dynamics \cite{sk97}. \\[2mm]
We would like to thank S. K\"ohler and M. Schr\"oder for useful discussions.  This work has been supported
by the Deutsche Forschungsgemeinschaft through SFB 410. 

\begin{figure}[tb]                              
\begin{minipage}[t]{7.1cm}\makebox[0cm]{}\\
        \psfig{file=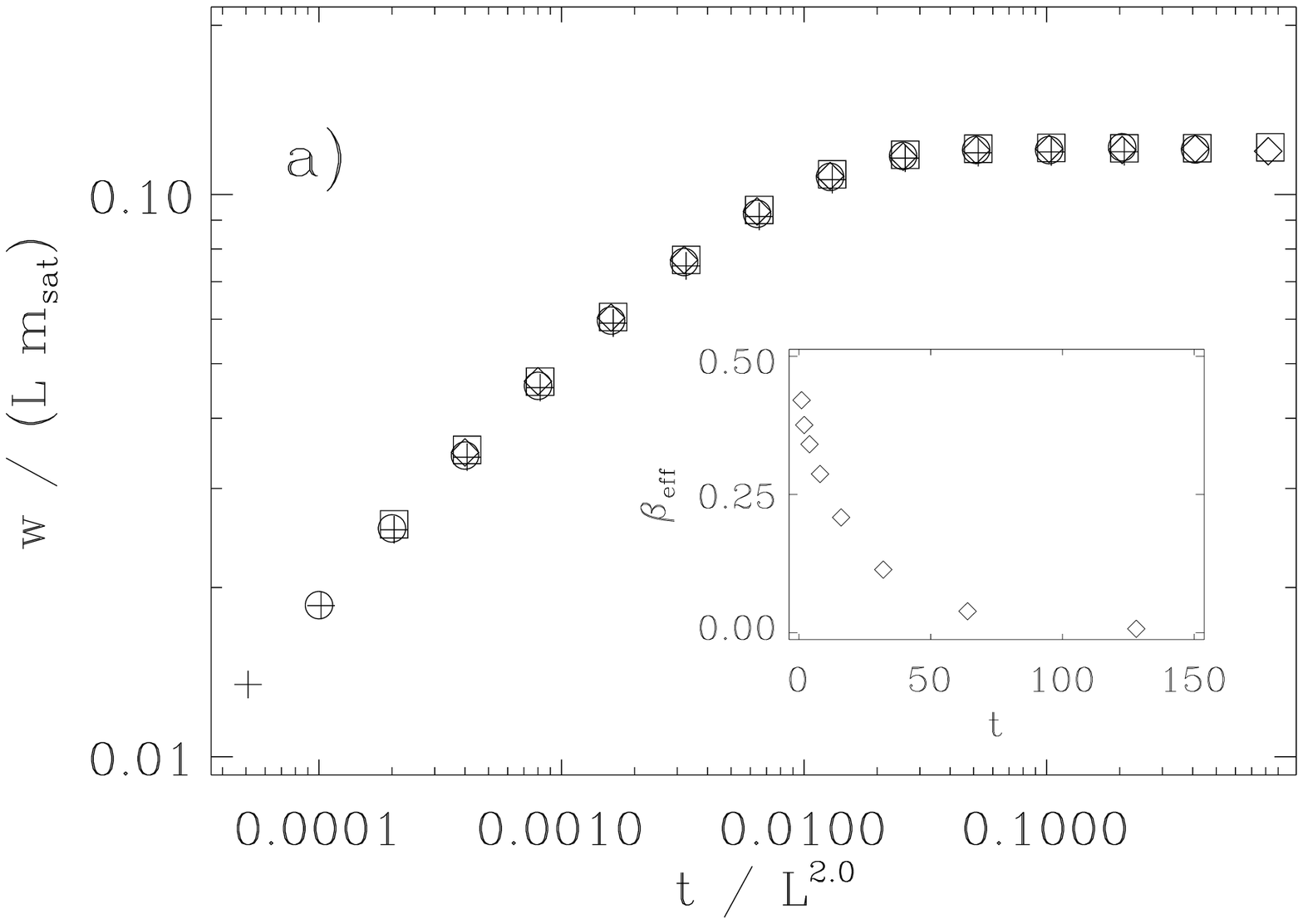,width=7.5cm}
\end{minipage} \hfill
\begin{minipage}[t]{7.1cm}\makebox[0cm]{}\\
        \psfig{file=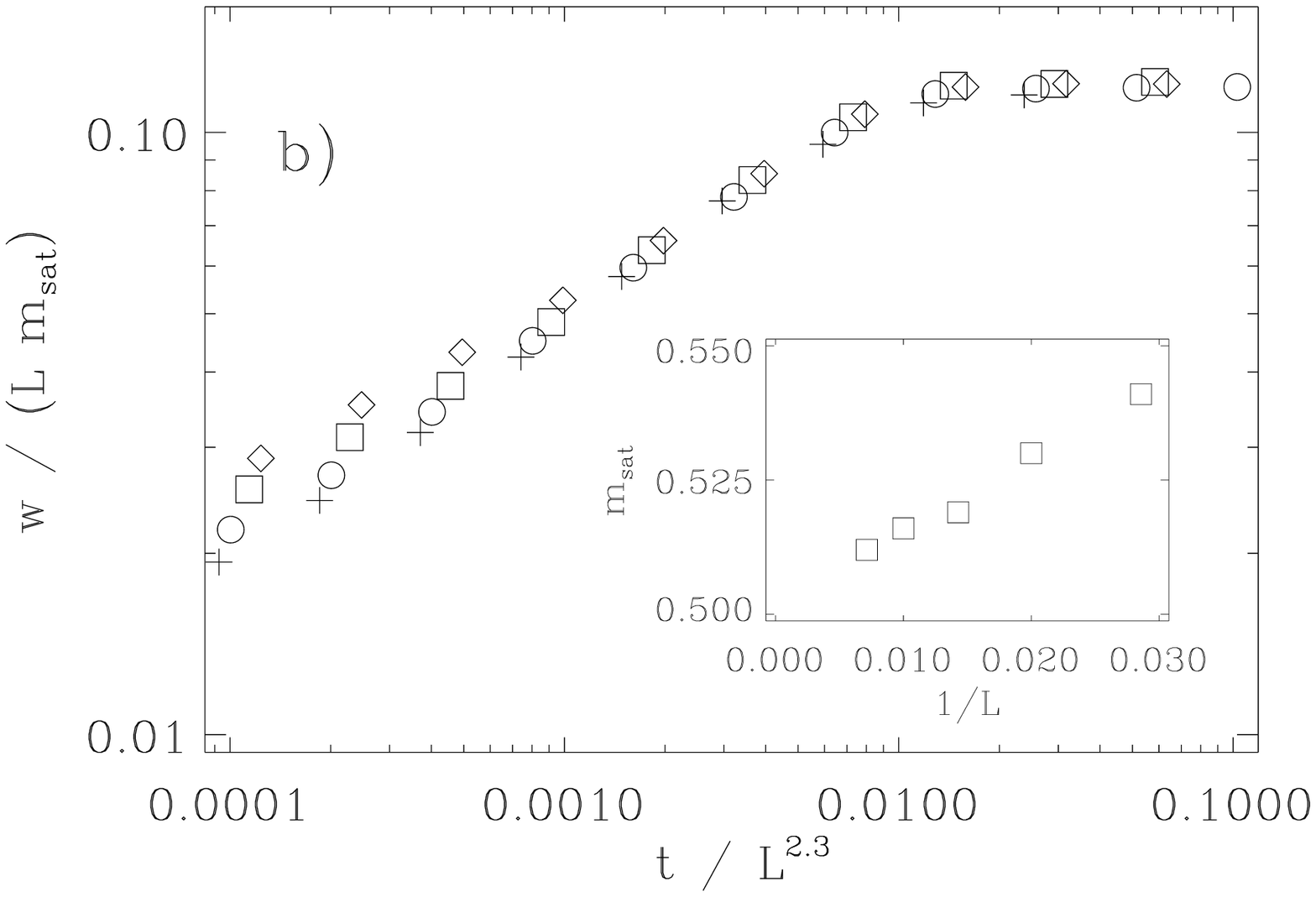,width=7.5cm}
\end{minipage}
\caption{  Scaling in 2+1 dimensions for system sizes $ L = 140 (+) $, $100 (\mbox{\large$\circ$})$,
  $ 70 (\Box)$, and $ 50 (\Diamond )$  with $l_{D2} = 2 L$.
 Results were averaged over 30 independent runs with error
bars smaller than the symbols. Fig.\  
a) shows the surface width vs.\ time for one island ($\protect{\sqrt{n_D}}\approx
L$), rescaled according to $ z=2.0 $ and $\alpha
=1$. The inset displays the effective exponent calculated from two
consecutive data points with $L=50$. In  
b) the surface width is plotted vs.\ time for the model with coarsening ($n_D = 200$), rescaled
according to $z=2.3$ and $\alpha=1$. The inset shows the saturation slope  vs.\ $1/L$. 
}
\end{figure}

\bibliography{Literatur}
\bibliographystyle{unsrt}

\end{document}